\documentclass[12pt]{iopart}

\pdfoutput=1

\usepackage{graphicx,epsfig,sidecap}
\usepackage{bm}
\usepackage{braket}
\usepackage{amsfonts,amssymb}

\usepackage{color}
\usepackage{times}
\usepackage[colorlinks,bookmarks=false,citecolor=blue,linkcolor=red,urlcolor=blue]{hyperref}

\def\be{\begin{equation}}
\def\ee{\end{equation}}

\def\bea{\begin{eqnarray}}
\def\eea{\end{eqnarray}}

\def\Tr{{\rm Tr}}

\def\l{\lambda}
\def\HH{\mathcal H}

\def\t#1{\overline{#1}}
\gdef\Db{{\overline\Delta}}
\newcommand{\tw}{{\cal T}}

\begin{document}

\title[Finite temperature entanglement negativity in CFT]
{Finite temperature entanglement negativity in conformal field theory}

\author{Pasquale Calabrese$^1$, John Cardy $^2$, and Erik Tonni$^3$}
\address{$^1$ Dipartimento di Fisica dell'Universit\`a di Pisa and INFN,
             Pisa, Italy.\\
          $^2$ Oxford University, Rudolf Peierls Centre for
          Theoretical Physics, 1 Keble Road, Oxford, OX1 3NP, United Kingdom
          and All Souls College, Oxford.\\
          $^3$ SISSA and INFN, via Bonomea 265, 34136 Trieste, Italy. }

\date{\today}

\begin{abstract}
We consider the logarithmic negativity of a finite interval embedded in an 
infinite one dimensional system at finite temperature. 
We focus on conformal invariant systems and we show that the naive approach based 
on the calculation of a two-point function of twist fields in a cylindrical geometry 
yields a wrong result. 
The correct result is obtained through a four-point function of twist fields
in which two auxiliary fields are inserted far away from the interval, 
and they are sent to infinity only after having taken the replica limit. 
In this way, we find a universal scaling form for the finite temperature negativity 
which depends on the full operator content of the theory and not only on the central charge.
In the limit of low and high temperatures, the expansion of this universal form can be obtained 
by means of the operator product expansion.  
We check our results against exact numerical computations for the critical harmonic chain.

\end{abstract}

\maketitle

\section{Introduction}

Entanglement is the key characteristic distinguishing quantum and classical mechanics. 
Since the birth of quantum mechanics this phenomenon has been attracting a continuous interest
in the physics community.
However, only during the last decade has a systematic study of the entanglement in extended 
quantum systems  been initiated (see e.g. Refs. \cite{rev} as reviews).
This study allowed a more precise characterisation of many-body quantum systems both in and out of equilibrium. 
In particular, it has been shown that when one considers the entanglement between two extended parts in an 
extended quantum system, universal scaling forms arise, e.g. close to a quantum critical 
point or in systems displaying topological order  (see again  Refs. \cite{rev} as reviews).

The entanglement between two extended parts in a {\it pure quantum state} is measured by the 
famous entanglement entropy, defined as follows.
Let $\rho$ be the density matrix of an extended quantum system taken 
in a pure quantum state $|\Psi\rangle$, so that $\rho=|\Psi\rangle\langle\Psi|$. 
Let us consider a spatial bipartition of the system in two parts $A$ and $B$ such that  
the Hilbert space can be written as a direct product $\HH=\HH_A\otimes\HH_B$. 
$A$'s reduced density matrix is $\rho_A=\Tr_B \rho$ and from this the (von Neumann) entanglement entropy is defined as
\be 
S_A= - {\rm Tr}\,\rho_A \ln \rho_A\,. 
\label{Sdef}
\ee
For a one-dimensional critical system whose scaling limit is described by a conformal
field theory (CFT), in the case when $A$ is an (extended) interval of length $\ell$ 
embedded in an infinite system, the asymptotic large $\ell$ behaviour of the entanglement 
entropy is  \cite{Holzhey,cc-04,Vidal,cc-rev}
\begin{equation}
\label{SA:asymp}
S_A=\frac{c}3 \ln \frac{\ell}a +{\rm const}\,,
\end{equation}
where $c$ is the central charge of the underlying CFT  \cite{c-lec} 
and $a$ the inverse of an ultraviolet cutoff (e.g. the lattice spacing).

One can then wonder if and how these nice universal results extend to finite temperature 
entanglement and whether it is possible by studying entanglement measures to describe 
the crossover/transition from quantum to classical world which is expected to take place as we ``warm up''
a quantum system.
Indeed this problem has been extensively studied (see again  Refs. \cite{rev} as reviews) at the level of few 
microscopic constituents (e.g. the entanglement between two spins in a very long spin-chain or the entanglement between 
few spins and the remainder of the system), but here we are genuinely interested in the 
entanglement between two extended parts at finite temperature, 
a problem which has been little studied \cite{AndersWinter,a-08,fcga-08} because of a series of technical difficulties. 
In this case, the entanglement entropy is no longer a good entanglement measure
since it clearly mixes up quantum and classical correlations (but still shows very interesting
features \cite{cc-04,ch-14,h-14}). 
A measure of entanglement for bipartite mixed states is instead the so-called entanglement negativity 
introduced in a seminal paper by Vidal and Werner \cite{vw-01} (see also \cite{v-98,ep-99,varmix}).
In order to define it, let us divide an extended quantum system in two parts 
which we call $A_1$ and $A_2$ (it is often very useful to think to a mixed state as 
obtained by tracing out the degrees of freedom of a larger system in a pure state which has been 
tripartite in  $A_1$, $A_2$ and $B$--a procedure called purification--). 
Let us denote by $|e_i^{(1)}\rangle$ and $|e_j^{(2)}\rangle$ two  bases in the Hilbert spaces  corresponding to 
$A_1$ and $A_2$ respectively. 
Let us define the partial transpose  with respect to $A_2$ degrees of freedom  as
\be 
\langle e_i^{(1)} e_j^{(2)}|\rho^{T_2}_{A_1\cup A_2}|e_k^{(1)} e_l^{(2)}\rangle=
\langle e_i^{(1)} e_l^{(2)}|\rho_{A_1\cup A_2}| e^{(1)}_k e^{(2)}_j\rangle,
\label{rhoAT2def}
\ee
and from this the  logarithmic negativity as
\be
{\cal E}\equiv\ln ||\rho^{T_2}_{A_1\cup A_2}||=\ln \Tr |\rho^{T_2}_{A_1\cup A_2}|\,,
\label{negdef}
\ee
where the trace norm  $||\rho^{T_2}_{A_1\cup A_2}||$ is
the sum of the absolute values of the eigenvalues of $\rho^{T_2}_{A_1\cup A_2}$.
In this expression $A_1\cup A_2$ does not have to coincide with the entire system. 
The negativity has the important property of being basis independent 
and this makes it calculable by means of quantum field theory (QFT) and in particular  CFT in which we are interested here. 
Recently \cite{us-letter,us-long} we developed a  systematic method to calculate the negativity  in
QFT and many-body systems in general.
This method has been so far extensively used to calculate the entanglement between two parts $A_1$ and $A_2$
of an extended system in a pure state in which a third part $B$ has been previously traced out. 
This method allowed the description of ground-states of CFTs \cite{us-letter,us-long,a-13,ctt-13,rr-14,kpp-14}, 
systems with topological order \cite{c-13,lv-13} and also some non-equilibrium situations \cite{ez-14,d-14,ctc-14}.
These studies also allowed a refined understanding of numerical and analytic 
computations \cite{Audenaert02,br-04,Neg1,Neg2,Neg3,sod2,kor1,kor2,sod1,sod4,sod3,cabcl-14}.
%


The manuscript is organised as follows. 
In Sec. \ref{Sec2} we discuss the general CFT approach to entanglement entropy and negativity. 
In Sec. \ref{Sec3} we report a first naive derivation of the negativity between a single interval and the 
rest of the system at temperature. This naive derivation provides a wrong result and we explain what is 
going wrong in this apparently innocuous computation. 
In Sec. \ref{Sec4} we fix the previous error and we provide a general scaling form for the negativity at 
finite temperature. 
Finally, in  Sec. \ref{Sec5} we report explicit numerical computation of the finite temperature negativity in 
the harmonic chain, showing perfect agreement with the previous universal result, but after all
the various lattice effects are properly taken into account. 
In Sec. \ref{Sec6} we draw our conclusions and discuss generalisations.

\section{General CFT approach to entanglement entropy and negativity}
\label{Sec2}

A very powerful way to calculate the entanglement entropy in a general QFT is based on a replica 
approach \cite{cc-04,cc-rev}. The overall idea is simply to compute the moments of the reduced 
density matrix $\Tr \rho_A^n$ with $n$ integer or, equivalently, the R\'enyi entropies
\be 
S^{(n)}_A= \frac{1}{1-n} \ln {\rm Tr}\,\rho_A^n\,. 
\label{Sndef}
\ee
The computation of these moments in 1+1 dimensional quantum field theories is made possible by the fact that  
they correspond to partition functions on $n$-sheeted Riemann surfaces with branch points at the boundaries between 
regions $A$ and $B$ \cite{cc-04}.
The partition function on the $n$-sheeted surface can be casted in the form of correlation functions on the complex plane
of particular operators termed branch-point twist fields \cite{cc-04,ccd-09,cc-rev}.
For example, in the case when $A=\cup_{i=1}^N [u_i,v_i]$ consists of $N$ disjoint intervals, we have
\be
\Tr \rho_A^n=
\langle \tw_n(u_1) \t\tw_n(v_1) \cdots\tw_n(u_N) \t\tw_n(v_N)  \rangle_{\mathbb C}\,.
\label{trrhongen}
\ee
At this point, if the expression for $\Tr \rho_A^n$ is analytically continuable to general complex $n$, 
the entanglement entropy is just given by the $n\to1$ limit of the R\'enyi entropies. 
Furthermore, the knowledge of the R\'enyi entropies for arbitrary $n$ provides 
also the full spectrum of the reduced density matrix \cite{cl-08}.

Although this approach has general validity in one spatial dimension, it is particularly useful in CFT 
because  the twist fields turned out to behave like primary operators with scaling dimension \cite{Knizhnik-87,cc-04}
\be
\Delta_{n}=\frac{c}{12}\Big(n-\frac1n\Big)\,. 
\ee
When this relation is specialised to an interval $A$ of length $\ell$ in an infinite system, we obtain the 
moments
\be
\Tr \rho_{A}^n= \langle {\cal T}_{n}(u) \overline{\cal T}_{n}(v)\rangle = c_n\left(\frac\ell{a} \right)^{-c/6(n-1/n)}\,,
\label{trnasy}
\ee
which can be straightforwardly analytically continued to get the von Neumann entropy (\ref{SA:asymp}).
The calculation of the entanglement entropy for many disjoint intervals $A=\cup_{i=1}^N [u_i,v_i]$
is much more complicated. 
Global conformal invariance fixes the scaling form
\be
\Tr \rho_{A}^n=
c_n^N\left({\prod_{j<k}(u_k-u_j)(v_k-v_j)\over\prod_{j, k}(v_k-u_j)}
\right)^{(c/6)(n-1/n)}{\cal F}_{n,N}(\{ x\}) \,.
\ee
Here $\{ x\}=\{ x_1,x_2\dots x_{2N-3}\}$ stands for the collection of $2N-3$ independent ratios that can be built with $2N$ points.
The functions ${\cal F}_{n,N}(\{ x\})$ are universal, but they depend on the full operator content of the CFT
--not only on the central charge-- and so 
they should be  calculated case by case.
Their calculation is a real tour de force which has been performed explicitly only in 
few instances \cite{cd-08,fps-08,cct-09,cct-11,c-10,atc-10,atc-11,h-10,rg-12,ctt-14} 
(exploiting old results of CFT on orbifolds \cite{cft-book,dixon,z-87,dvv-87}) 
and agree with some explicit numerical and analytical computations 
\cite{cd-08,fps-08,atc-10,atc-11,ch-09,hol,ip-09,fc-10,fc-11,f-12,ctt-14}.
We mention that these universal functions ${\cal F}_{n,N}(\{ x\})$ have very cumbersome analytic forms which 
prevented the analytic continuations to $n\to1$, with the notable exception of the free un-compactified boson 
\cite{cct-09} which is the only exactly known entanglement entropy for two disjoint intervals. 
Some results for two disjoint regions in higher dimensional CFTs are also known \cite{c-13b,s-12,s-14}.

\subsection{The entanglement negativity in CFT}

Moving now to the entanglement negativity, 
we start by reviewing the general situation studied by CFT in Refs. \cite{us-letter,us-long} in which a one-dimensional system 
(which can be either infinite, finite, semi-infinite etc.)
is divided into three parts, two of which are finite intervals $A_1$ and $A_2$ of length 
respectively $\ell_1$ and $\ell_2$ and the part $B$ represents the remainder of the system. 
We will denote with $\rho_A$ the reduced density matrix of $A\equiv A_1\cup A_2=[u_1,v_1]\cup [u_2,v_2]$, 
i.e. $\rho_A=\rho_{A_1\cup A_2}$, which is obtained by tracing out the part $B$ of the system, i.e. $\rho_A=\Tr_B \rho$.  

The quantum field theory approach to  negativity  is also based on a replica trick \cite{us-letter,us-long}:  
one considers the traces $\Tr (\rho_A^{T_2})^n$ of integer powers of $\rho_A^{T_2}$ 
and notices that $\Tr (\rho_A^{T_2})^n$ have a different functional dependence on $|\lambda_i|$ 
according to the parity of $n$ ($\lambda_i$ being the eigenvalues of $\rho_A^{T_2}$).
Indeed,  for $n$ even and odd (that we denote as $n_e$ and $n_o$ respectively), 
the traces of integer powers of $\rho_A^{T_2}$ are
\bea
\Tr (\rho_A^{T_2})^{n_e}&=&\sum_i \lambda_i^{n_e}= \sum_{\l_i>0} |\l_i|^{n_e}+ \sum_{\l_i<0} |\l_i|^{n_e}\,, 
\label{trne}\\
\Tr (\rho_A^{T_2})^{n_o}&=&\sum_i \lambda_i^{n_o}= \sum_{\l_i>0} |\l_i|^{n_o}- \sum_{\l_i<0} |\l_i|^{n_o}\,.
\label{trno}
\eea
If we set $n_e=1$ in Eq. (\ref{trne}) we formally obtain $ \Tr |\rho_A^{T_2}|$, which is the quantity we are 
interested in, according to Eq. (\ref{negdef}). 
Instead, setting $n_o=1$ in Eq. (\ref{trno}) gives the normalization $\Tr \rho_A^{T_2}=1$.
This means that the analytic continuations from even and odd $n$ are different and, in particular, 
the trace norm that we want to compute is obtained
by considering the analytic continuation of the even sequence at $n_e\to1$, 
i.e. 
\be  {\cal E}=\lim_{n_e\to1} \ln \Tr (\rho_A^{T_2})^{n_e}\,.
\ee

Since in the QFT representation of the density matrix a transposition has the effect to 
exchange lower and upper edges of the open cut \cite{us-letter,us-long} (see also next section),
the traces of integer powers of the partial transpose for two disjoint intervals  
are partition functions on $n$-sheeted Riemann surfaces or, equivalently, 
the correlation functions of four twist fields \cite{us-long}
\be
\Tr(\rho_A^{T_{2}})^n=
\langle {\cal T}_n(u_1)\overline{\cal T}_n(v_1) \overline{\cal T}_n(u_2){\cal T}_n(v_2)\rangle\,,
\label{4ptdef}
\ee
i.e. the partial transposition has the net effect of exchanging two twist operators compared to 
$\Tr\rho_A^n$, cf. Eq. (\ref{trrhongen}) for $N=2$.

Eq. (\ref{4ptdef}) is generically valid, but it simplifies when specialised to the case of two adjacent intervals, 
obtained by letting $v_1\to u_2$,  leading to the three-point function
\be
\Tr(\rho_A^{T_{2}})^n=
\langle {\cal T}_n(u_1) \overline{\cal T}_n^2(u_2){\cal T}_n(v_2)\rangle\,.
\label{3ptdef}
\ee

A further simplification occurs when specialising to a {\it pure state} by letting $B\to \emptyset$ 
(i.e. $u_2\to v_1$ and $v_2\to u_1$, when dealing with the CFT on the plane) for which 
$\Tr (\rho_A^{T_2})^{n}$ becomes a two-point function 
\be
\Tr (\rho_A^{T_2})^{n}=\langle {\cal T}^2_{n}(u_2) \overline{\cal T}^2_{n}(v_2)\rangle\,.
\ee
As explained in more details in Ref.~\cite{us-long}, as a partition function on an $n$-sheeted 
Riemann surface, this expression depends on the parity of $n$ because  
${\cal T}_n^2$ connects the $j$-th sheet with the $(j+2)$-th one. 
For $n=n_e$ even, 
the $n_e$-sheeted Riemann surface decouples in two independent ($n_e/2$)-sheeted 
surfaces.
Conversely for $n=n_o$ odd, the surface remains a $n_o$-sheeted Riemann
surface. In formulas, these observations are
\bea
\Tr (\rho_A^{T_2})^{n_e}&=& (\langle {\cal T}_{n_e/2}(u_2) \overline{\cal T}_{n_e/2}(v_2)\rangle)^2=
(\Tr\rho_{A_2}^{n_e/2})^2\,,
\label{pureqfte} \\
\Tr (\rho_A^{T_2})^{n_o}&= & \langle {\cal T}_{n_o}(u_2) \overline{\cal T}_{n_o}(v_2)\rangle=\Tr \rho_{A_2}^{n_o}\,.
\label{pureqft}
\eea
Hence, for a bipartite system, $\Tr (\rho_A^{T_2})^{n}$ can be generically written as a function of  $\Tr \rho_{A_2}^m$, 
with $m$ being either $n$ or $n/2$. 
In particular, taking the limit $n_e\to1$, we obtain that the logarithmic negativity 
equals the R\'enyi entropy of order $1/2$, a well known result for bipartite states \cite{vw-01}.


Eqs. (\ref{pureqfte}) and (\ref{pureqft}) are simply specialised to the CFT case by plugging into them 
the general formula for $\Tr \rho_{A_2}^n$ in Eq. (\ref{trnasy}), obtaining  \cite{us-letter,us-long}
\be\fl
\Tr (\rho_A^{T_2})^{n_e}= (\langle {\cal T}_{n_e/2}(u_2) \overline{\cal T}_{n_e/2}(v_2)\rangle)^2=(\Tr\rho_{A_2}^{n_e/2})^2
=c_{n_e/2}^2 \Big(\frac{\ell}a\Big)^{-{c}/{3}({n_e}/2-2/{n_e})}\,,
\label{1inte}
\ee
and 
\be\fl
\Tr (\rho_A^{T_2})^{n_o}=  \langle {\cal T}_{n_o}(u_2) \overline{\cal T}_{n_o}(v_2)\rangle=\Tr \rho_{A_2}^{n_o}=
c_{n_o} \Big(\frac{\ell}a\Big)^{-{c}/6(n_o-1/{n_o})}\,.
\label{1into}
\ee
This simple result shows an important feature of the negativity in CFT, i.e.
the operator ${\cal T}^2_{n_e}$ and $\overline{\cal T}^2_{n_e}$ have scaling dimension $\Delta^{(2)}_n $ which depends
on the parity of the replica index as 
\be
\Delta^{(2)}_n \equiv  
\left\{ \begin{array}{ll}
\displaystyle
\Delta_n
\hspace{.5cm}& 
\textrm{odd $n$}
\\
\displaystyle
2\Delta_{n/2}
\hspace{.5cm}& 
\textrm{even $n$}
\end{array}
\right. .
\ee
The constants $c_{n_o,n_e}$ are non-universal, but they are the same 
as in the R\'enyi entanglement entropies (\ref{trnasy}), so that we can write
\be
\Tr (\rho_A^{T_2})^{n}=c_{n}^{(2)} \Big(\frac{\ell}a\Big)^{-\Delta_n^{(2)}},
\qquad 
c_{n}^{(2)}\equiv  
\left\{ \begin{array}{ll}
\displaystyle
c_n
\hspace{.5cm}& 
\textrm{odd $n$}
\\
\displaystyle
c_{n/2}^2
\hspace{.5cm}& 
\textrm{even $n$}
\end{array}
\right. .
\label{cn2def}
\ee

In the replica limit $n_e \rightarrow 1$, we have $\Delta_{n_e} \rightarrow 0$ and $\Delta^{(2)}_{n_e} \rightarrow -c/4$.
Thus, performing the analytic continuation, we have 
\be\fl
|| \rho_A^{T_2}||= \lim_{n_e\to1} \Tr (\rho_A^{T_2})^{n_e}=
c_{1/2}^2  \Big(\frac{\ell}a\Big)^{{c}/2}
\qquad  \Rightarrow \qquad
{\cal E}=\frac{c}2\ln \frac{\ell}a+2\ln c_{1/2}\,,
\label{neg2pt}
\ee
which again is the result that for pure bipartite states the logarithmic negativity equals the 
R\'enyi entropy of order $1/2$.

\section{A naive finite temperature result for the negativity of a single interval in infinite space and why it goes wrong}
\label{Sec3}

In a general one-dimensional quantum field theory, finite temperature expectation values
are simply obtained by evaluating the path integral on a cylinder of circumference $\beta=1/T$
in the imaginary time direction.  
This correspondence is particularly useful in a CFT, because a cylindrical geometry can be 
obtained from the complex plane by the conformal transformation
\be
z\to w=(\beta/2\pi)\ln z,
\ee 
where $z$ is the complex coordinate on the plane and $w=\sigma + {\rm i} \tau$ lives on the cylinder. 
Under the conformal mapping $z\to w=w(z)$, the correlation functions of primary operators transform like \cite{cft-book}
\be
\langle\Phi_1(z_1,\bar z_1)\Phi_2(z_2,\bar z_2)\ldots\rangle
=\prod_j|w'(z_j)|^{\Delta_j}
\langle\Phi_1(w_1,\bar w_1)\Phi_2(w_2,\bar w_2)\ldots\rangle\,.
\label{genmap}
\ee

Then the obvious ``temptation" is to apply the above formula to the correlation function of twist fields determining the 
negativity of an interval of length $\ell$ in an infinite system at temperature $1/\beta$, i.e. 
assuming 
\be
\Tr (\rho^{T_A})^{n}=\langle {\cal T}^2_{n}(u) \overline{\cal T}^2_{n}(v)\rangle_\beta\,,
\label{wrong}
\ee
where $\langle \cdot\rangle_\beta$ stands for expectation values on the cylinder. 
We anticipate that this apparently innocuous formula is wrong.
Notice that, since from now on we will consider only bipartite systems, we denoted the finite interval with $A$ and 
the remainder of  the system with $B$, but compared to the previous sections we change notations 
as $(A_1,A_2,B)\to (A,B,\emptyset)$.
 
The rhs of Eq. (\ref{wrong}) can be derived using Eq. (\ref{genmap}), providing   for even $n$
\be
\langle {\cal T}^2_{n_e} (0)\overline{\cal T}^2_{n_e}(\ell)\rangle_{\beta}=
c_{n_e/2}^2\left(\frac{\beta}{\pi a} \sinh\frac{\pi \ell}{\beta}\right)^{-c/3(n_e/2-2/n_e)}\,,
\ee
that analytically continued to $n_e=1$ gives
\be\fl
|| \rho^{T_A}||= c_{1/2}^2\left(\frac{\beta}{\pi a} \sinh\frac{\pi \ell}{\beta}\right)^{c/2},\qquad 
\Rightarrow\;\; {\cal E}_{\rm naive}=\frac{c}2\ln \left(\frac{\beta}{\pi a} \sinh\frac{\pi \ell}{\beta}\right)+2\ln c_{1/2}\,.
\label{naive}
\ee
This result resembles the one for the entanglement entropy at finite temperature \cite{cc-04}, 
and indeed it is nothing but the finite temperature R\'enyi entropy of order $1/2$.
But this naive result makes no sense at all for several reasons. First, for the finite temperature  
mixed state, the R\'enyi entropy is {\it not} an entanglement measure while the negativity is.
However, one could pretend that for CFTs the two could coincide, but this is not the case. 
Indeed, let us look at this result for fixed $\ell$ and increasing the temperature from $0$ to some finite value $T=1/\beta$. 
Eq. (\ref{naive}) is a monotonous increasing function of $T$, 
which goes against the common sense  suggesting the entanglement to be
a decreasing function of the temperature. 
Even more dramatically for very high temperature, when the system becomes classical Eq. (\ref{naive}) predicts an infinite 
entanglement, which is absurd for a classical system.

\begin{figure}[t]
\vspace{.4cm}
\begin{center}
\includegraphics[width=.5\textwidth]{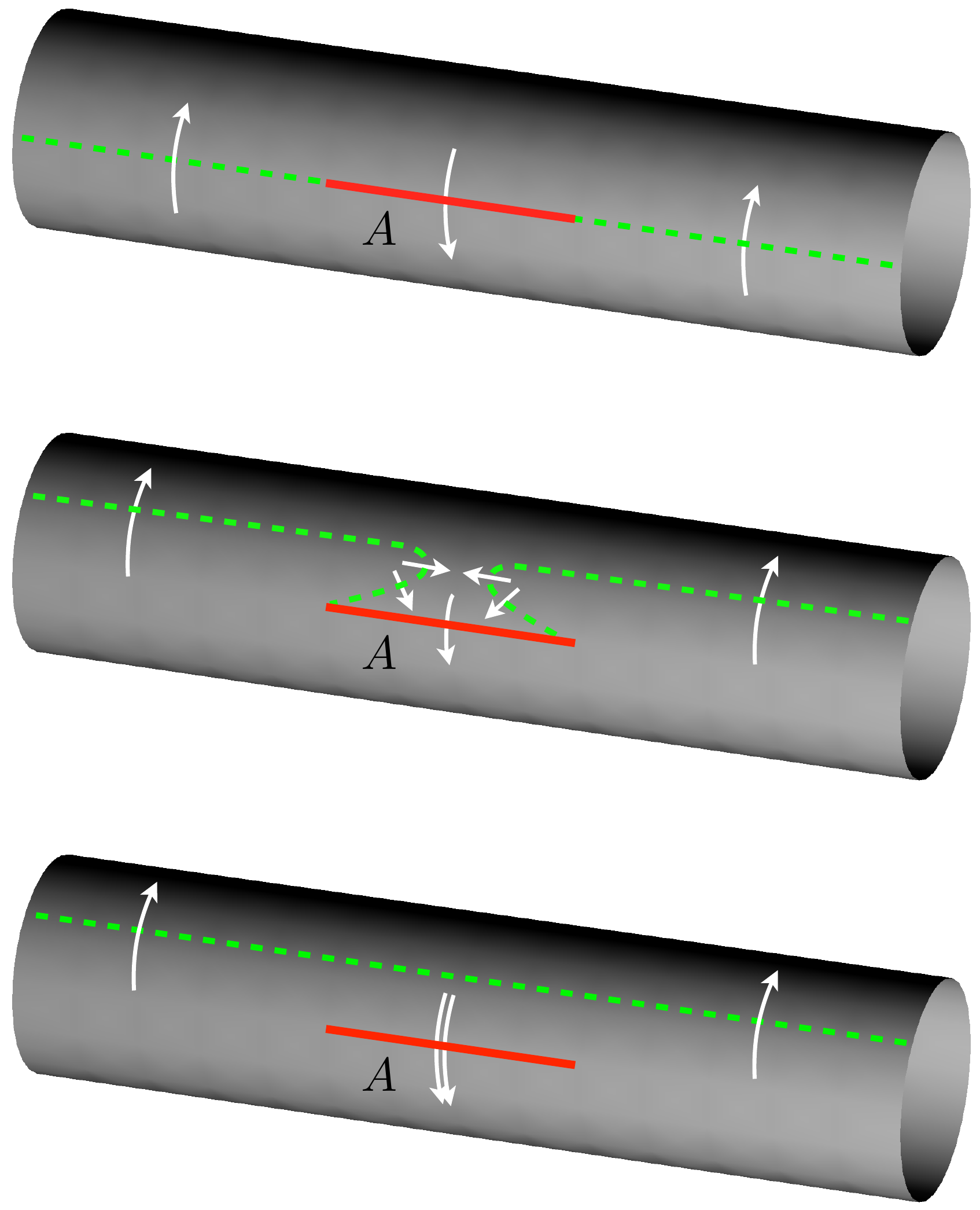}
\end{center}
\vspace{-.1cm}
\caption{Partial transposition and $\Tr (\rho^{T_A})^n$ of one interval $A$ at finite temperature.
Top: 
simple arrows indicate that in  $\Tr (\rho^{T_A})^n$ one passes from the $j$-th copy to the $(j+1)$-th 
one by following them through the cut (depending on the verse of the arrow).
Middle:
deforming the dashed green line as indicated, part of it becomes a cut extending along the whole cylinder and parallel to its axis, while the remaining part merges with the red segment.
Bottom:
The double arrow denotes a double jump, from the $j$-th copy to the $(j+2)$-th one, following it through the cut. 
The $j$-th copy is sewed to the $(j+1)$-th one through the dashed line.  
Because of this connection, the $n_e$-sheeted Riemann surface occurring in $\Tr (\rho^{T_A} )^{n_e}$ does not factorise 
into the product of two identical $(n_e/2)$-sheeted Riemann surfaces.
}
\label{fig_argument}
\end{figure}

Let us now try to understand what is going wrong in the apparently simple and straightforward calculation above. 
In order to do so we should first go back to the definition of density matrix in a finite temperature QFT. 
This is given by a cylinder of circumference $\beta$ with an open cut on the axis $\tau=0$.
The upper and lower edges of the cut are the rows and the columns of the density matrix. 
Then matrix multiplication corresponds to sew together an upper and lower edge of two cylinders.
To give a practical example, $\Tr \rho^n$ is nothing but the partition function of $n$ of those cylinder joined cyclically 
in such a way to have a ``fatter'' cylinder of circumference $n\beta$. 
As explained in more details in Ref. \cite{cc-04}, in the reduced density matrix of a segment $A$ one should sew the cut 
along $B$ inside the single cylinder, in such a way that when constructing $\Tr \rho_A^n$ 
we end up in a $n$-sheeted Riemann surface in which the cylinders are connected cyclically only along the interval $A$.

The construction of the partial transpose has been derived in Refs. \cite{us-letter,us-long} to which we refer for more details.
The partial transposition with respect to an interval $A$ amounts to exchange rows and columns corresponding to that interval, 
which is exchanging the lower and upper edges of the cylinder along $A$ in the QFT construction.
Thus, when we consider $\Tr (\rho^{T_A})^n$, the resulting surface is such that the axis at $\tau=0$ on the $j$-th cylinder 
is connected to the upper cylinder along $B$ but to the lower one along $A$. 
In the top picture of Fig. \ref{fig_argument}, this is shown with arrows of different verses, pointing up and down to respectively 
indicate the connections to the $(j+1)$-th cylinder when moving following the arrow.
We can now deform the cut associated to $B$ as shown in the middle picture of Fig. \ref{fig_argument} 
and this deformation does not change the topology of the surface. 
Proceeding with the deformation, the $j$-th sheet becomes the cylinder depicted in the bottom picture: 
it is connected to the $(j+1)$-th cylinder through the dashed green line and to the $(j+2)$-th one through the interval $A$
(represented pictorially as a double arrow). 
There is no way to remove this dashed green line which connects consecutive sheets and therefore
for $n=n_e$ the corresponding Riemann surface does not decouple into two identical $(n_e/2)$-sheeted Riemann surfaces. 
Always because of such connection through the dashed green line, the Riemann surface occurring in   
$\Tr (\rho^{T_A} )^{n_o}$ is not the same one of $\Tr \rho_A^{n_o}$. 
Thus, at finite temperature $\Tr (\rho^{T_A} )^{n}$ cannot be related to $\Tr \rho_A^{n}$, as done in \cite{us-letter,us-long} for $T=0$.

\begin{figure}[t]
\vspace{.4cm}
\begin{center}
\includegraphics[width=1\textwidth]{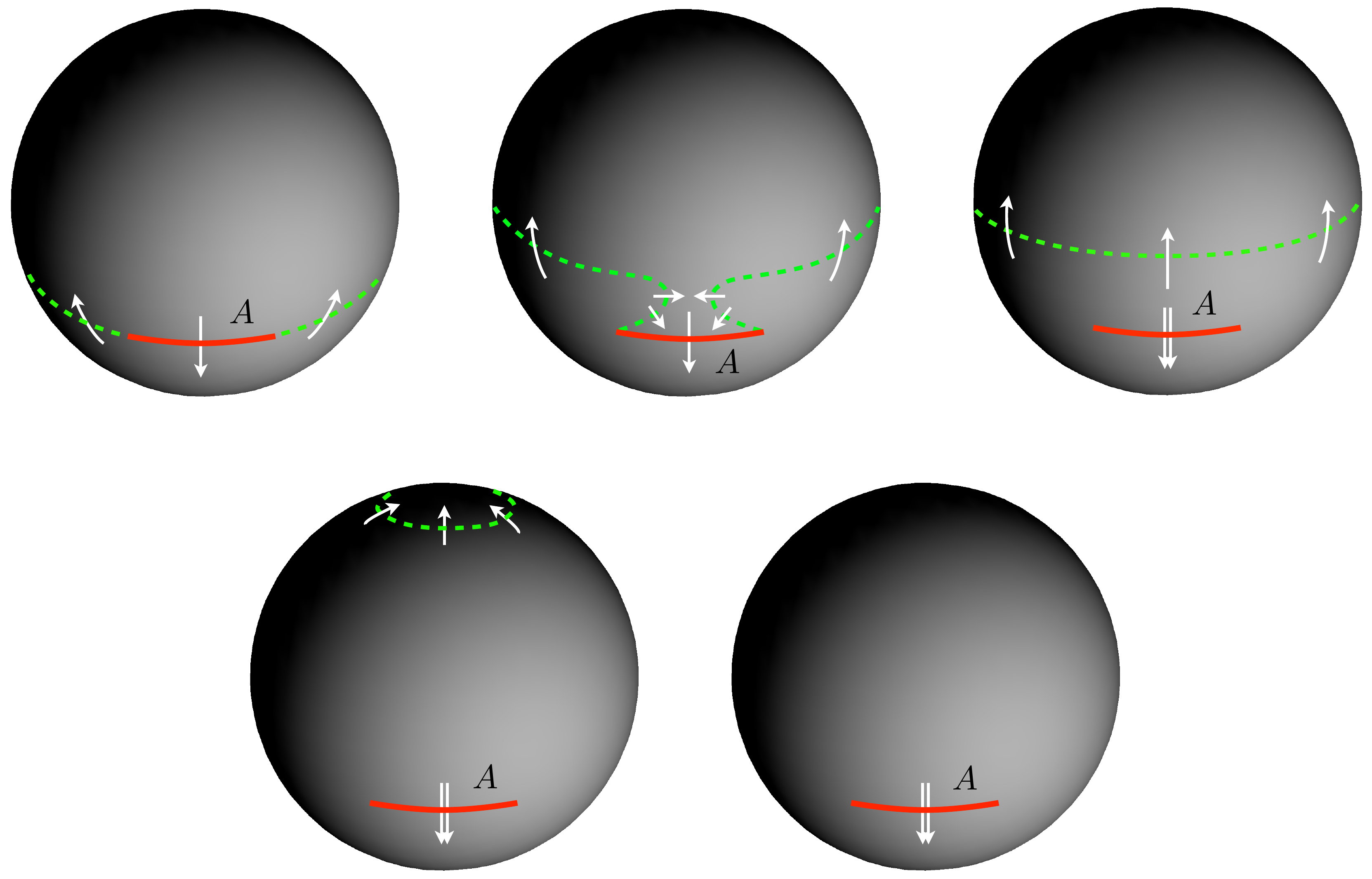}
\end{center}
\vspace{-.1cm}
\caption{The deformation procedure described for Fig. \ref{fig_argument} (with the same notations) performed at zero temperature, 
i.e. on the sphere.
In this case the cut given by the dashed green line can be shrunk to a point and it annihilates, leaving only the cut through $A$, 
which connect the $j$-th copy to the $(j+2)$-th one.
Thus, at zero temperature, the $n_e$-sheeted Riemann surface occurring in $\Tr (\rho^{T_A} )^{n_e}$ factorizes into the product 
of two identical $(n_e/2)$-sheeted Riemann surfaces.
}
\label{fig_argument sphere}
\end{figure}

At this point it is natural to wonder why a similar problem does not arise at $T=0$ in which case the cylinder degenerates to the 
whole complex plane which is topologically equivalent to a sphere. 
In Fig. \ref{fig_argument sphere} we proceed in the same way as in Fig. \ref{fig_argument}, but substituting the
cylinder with a sphere. In the first three panels everything proceeds in complete analogy to the cylinder case 
resulting finally in a dashed green line parallel to the equator. 
However, in this case, the dashed green line can be shrunk to a point and therefore it vanishes.
Thus, since the $j$-th sheet is connected only to the $(j+2)$-th one along the segment $A$, we are allowed to use the 
construction in Ref.  \cite{us-long} to relate, at zero temperature, $\Tr (\rho^{T_A} )^{n}$ to $\Tr \rho_A^{n}$.

A final comment is necessary at this point. The above reasoning shows that the negativity cannot be calculated 
by a simple mapping from the cylinder to the sphere if the partial transposition involves an infinite part  
of an infinite system at finite temperature. 
This is not the case if one, for example, is interested in the negativity between two (adjacent or disjoint) 
finite intervals at finite temperature. 
For this reason, to quote a concrete example, the finite temperature calculation of the negativity between two adjacent intervals in 
Ref. \cite{ez-14}, which is based on the calculation of a three-point function on the plane, is indeed free from these troubles and 
therefore correct.

\section{The correct finite temperature negativity for one interval in the infinite line}
\label{Sec4}

After having understood what goes wrong in Eq. (\ref{wrong}), we should find a way to take into account the 
branch cut (i.e. the dashed-green line in Fig. \ref{fig_argument}) in the partial transposed reduced density matrix. 
Denoting by $w = \sigma +\textrm{i} \tau$ ($\sigma \in \mathbb{R}$ and $\tau \in [0,\beta)$) the complex coordinate on the cylinder,
the easiest way to tackle the problem is to let the branch cut starting and ending not at ${\rm Re} [w]=\pm \infty$,
but at some large and finite values that we denote as $w=\pm L$.
Proceeding according to this idea, we should then calculate 
the following four-point function on the infinite cylinder of circumference $\beta$
\be
\langle 
\mathcal{T}_n(-L)  \overline{\mathcal{T}}^2_n(-\ell)  \mathcal{T}^2_n(0)  \overline{\mathcal{T}}_n(L) \rangle_\beta \,,
\label{4point cylinder}
\ee
where $L > \ell>0$.
The locations of the twist operators are aligned along the line $\tau=0$ parallel to the axis of the cylinder. 
In order to compute the negativity, we should first calculate the four-point function above for finite $L$,
then take the replica limit $n_e\to1$ and only after the limit $L\to\infty$. 
As we will see explicitly below, the two limits do not commute explaining why we get a different result 
compared to the naive one (\ref{naive}).
For simplicity in the formulas, we will omit the dependence on the inverse UV cutoff $a$ during the 
course of the calculation and we will restore it only at the end. 
Notice that the above prescription provides automatically that the negativity of part $B$ of the system 
${\cal E}_B$ equals ${\cal E}_A$, because in Eq. (\ref{4point cylinder}) the exchange $A$ and $B$
is obtained by interchanging ${\cal T}_n \leftrightarrow \overline{\cal T}_n$ and ${\cal T}^2_n \leftrightarrow \overline{\cal T}_n^2$.

By employing the conformal map $z=e^{2\pi w/\beta}$ and using Eq. (\ref{genmap}), we can write Eq. 
(\ref{4point cylinder}) in terms of the corresponding four-point function on the complex plane, namely
\bea
\label{4point cylinder-sphere}
\langle 
\mathcal{T}_n(-L)  \overline{\mathcal{T}}^2_n(-\ell)  \mathcal{T}^2_n(0)  \overline\mathcal{T}_n(L) 
\rangle_\beta= \nonumber\\
\qquad =
\bigg(\frac{2\pi}{\beta}\bigg)^{2\Delta_n+2\Delta_n^{(2)}}
\frac{\langle 
\mathcal{T}_n(e^{-2\pi L/\beta})  \overline{\mathcal{T}}^2_n(e^{-2\pi \ell/\beta})  
\mathcal{T}^2_n(1)  \overline\mathcal{T}_n(e^{2\pi L/\beta}) 
\rangle_{\mathbb C}}{
e^{2\pi \Delta^{(2)}_n \ell/\beta}
}\,.
\eea
On the complex plane, a general four-point function of primary fields $V_j(z_j, \bar{z}_j)$ with different scaling dimensions $\Delta_{(j)}$ 
can be written, for instance, as follows 
\be\fl
\label{4points corr sphere}
\langle 
\prod_{j=1}^4
V_j(z_j, \bar{z}_j)
\rangle_{\mathbb C}
=
\frac{z_{13}^{\Delta_{(1)} + \Delta_{(3)} }  z_{24}^{\Delta_{(2)} + \Delta_{(4)} } }{
z_{12}^{\Delta_{(1)} + \Delta_{(2)} }  z_{23}^{\Delta_{(2)} + \Delta_{(3)} } 
z_{34}^{\Delta_{(3)} + \Delta_{(4)} }  z_{14}^{\Delta_{(1)} + \Delta_{(4)} } 
}\, 
\mathcal{G}(x),
\qquad
x\equiv \frac{z_{12} z_{34}}{z_{13} z_{24}},
\ee
being $z_{ij}\equiv |z_i-z_j|$ and $x$ the cross ratio of the four points. 
Specifying Eq. (\ref{4points corr sphere}) to our case, the four-point function occurring 
in the rhs of Eq. (\ref{4point cylinder-sphere}) can be written as 
\be
\label{4point-func twist}
\langle 
\mathcal{T}_n(z_1)  \overline{\mathcal{T}}^2_n(z_2)  
\mathcal{T}^2_n(z_3)  \overline{\mathcal{T}}_n(z_4) 
\rangle
=
\frac{\mathcal{G}_n(x)}{z_{14}^{2\Delta_n}  z_{23}^{2\Delta^{(2)}_n} x^{\Delta_n+ \Delta^{(2)}_n}},
\ee
where the points $z_i$ are still arbitrary complex numbers.
It is clear that there is a degree of arbitrariness in the way of writing the above  four-point function since 
any power of the harmonic ratio $x$ can be absorbed in the scaling function ${\cal G}_n(x)$.
We will use this arbitrariness to our advantage in such a way to have reasonable limits for $x\to 0$ and $x\to1$.
To fix these limits we must consider the OPEs
\be\fl 
\label{OPE id}
\mathcal{T}_n(u)  \overline{\mathcal{T}}_n(v)  
= 
\frac{c_n}{|u-v|^{2\Delta_n}} \, \mathbb{I}+ \dots,
\qquad
\mathcal{T}^2_n(u)  \overline{\mathcal{T}}^2_n(v)  
= 
\frac{c_n^{(2)}}{|u-v|^{2\Delta_n^{(2)}}} \, \mathbb{I}+ \dots,
\quad
u \rightarrow v,
\ee
and
\be
\label{OPE 2}
\mathcal{T}_n(u)  \overline{\mathcal{T}}^2_n(v)  
= 
\frac{C_{\mathcal{T}_n \overline{\mathcal{T}}_n^2 \overline{\mathcal{T}}_n }}{|u-v|^{\Delta_n^{(2)}}} \, 
\overline{\mathcal{T}}_n(v) + \dots,
\qquad
u \rightarrow v,
\ee
with the corresponding OPE for $\overline{\mathcal{T}}_n(u)  \mathcal{T}^2_n(v) $ 
with $C_{\mathcal{T}_n \overline{\mathcal{T}}_n^2 \overline{\mathcal{T}}_n }=C_{\overline{\mathcal{T}}_n \mathcal{T}_n^2 \mathcal{T}_n }$.

From Eq. (\ref{OPE id}), we have the following cluster property for $z_3 \rightarrow z_2$ and $z_4 \rightarrow z_1$:
\be\fl
\label{expansion x=1}
\langle 
\mathcal{T}_n(z_1)  \overline{\mathcal{T}}^2_n(z_2)  
\mathcal{T}^2_n(z_3)  \overline{\mathcal{T}}_n(z_4) 
\rangle
=
\langle 
\mathcal{T}_n(z_1)  \overline{\mathcal{T}}_n(z_4)  
\rangle\,
\langle 
 \overline{\mathcal{T}}^2_n(z_2)   \mathcal{T}^2_n(z_3)
\rangle
 +\dots,
 \ee
while from Eq. (\ref{OPE 2}) we have
\be\fl 
\label{expansion x=0}
\langle 
\mathcal{T}_n(z_1)  \overline{\mathcal{T}}^2_n(z_2)  
\mathcal{T}^2_n(z_3)  \overline{\mathcal{T}}_n(z_4) 
\rangle
=
\frac{
C_{\mathcal{T}_n \overline{\mathcal{T}}_n^2 \overline{\mathcal{T}}_n }^2
c_n
}{
(z_{12} \,z_{34})^{\Delta^{(2)}_n} z_{13}^{2\Delta_n}}
+\dots,
\qquad
z_2 \rightarrow z_1 ,
\hspace{.4cm}
z_4 \rightarrow z_3,
\ee
where the dots denote less divergent terms.
The expansions (\ref{expansion x=1}) and (\ref{expansion x=0}) correspond respectively to $x\rightarrow 1$ and $x\rightarrow 0$. 
These expressions suggest us to write the four-point function of twist fields in the following form
\be
\label{4-point function ok}
\langle 
\mathcal{T}_n(z_1)  \overline{\mathcal{T}}^2_n(z_2)  
\mathcal{T}^2_n(z_3)  \overline{\mathcal{T}}_n(z_4) 
\rangle
=
\frac{c_n \, c_n^{(2)}}{z_{14}^{2\Delta_n}  z_{23}^{2\Delta^{(2)}_n}}
\frac{\mathcal{F}_n(x)}{x^{\Delta^{(2)}_n}},
\ee
where (\ref{expansion x=1}) and (\ref{expansion x=0}) impose respectively that
\be
\label{bc on F}
\mathcal{F}_n(1) =1,
\qquad
\mathcal{F}_n(0) =\frac{C_{\mathcal{T}_n \overline{\mathcal{T}}_n^2 \overline{\mathcal{T}}_n }^2}{c_n^{(2)}}.
\ee
This choice of the universal function ${\cal F}_n(x)$ is such that it has finite limits both for $x\to0$ and $x\to1$.

Specialising to the four points in the rhs of Eq. (\ref{4point cylinder-sphere}) we have
\bea\fl
\langle 
\mathcal{T}_n(e^{-2\pi L/\beta})  \overline{\mathcal{T}}^2_n(e^{-2\pi \ell/\beta})  
\mathcal{T}^2_n(1)  \overline\mathcal{T}_n(e^{2\pi L/\beta}) 
\rangle_{\mathbb C}
=\nonumber\\ \hspace{2cm}
= \frac{c_n \, c_n^{(2)}}{(2\sinh (2\pi L/\beta))^{2\Delta_n}  (1- e^{-2\pi \ell/\beta})^{2\Delta^{(2)}_n}}
\frac{\mathcal{F}_n(x)}{x^{\Delta^{(2)}_n}},\label{interme}
\eea
and the harmonic ratio becomes 
\be
x=\frac{(e^{-2\pi \ell/\beta}-e^{-2\pi L/\beta})(e^{2\pi L/\beta}-1)}{
(1- e^{-2\pi L/\beta})(e^{2\pi L/\beta}-e^{-2\pi \ell/\beta})} \,,
\ee
whose large $L$ limit is 
\be
\lim_{L\to\infty}x = e^{-2\pi \ell/\beta}\,.
\ee
Plugging Eq. (\ref{interme}) in Eq. (\ref{4point cylinder-sphere}), the desired cylinder four-point function is
\bea\fl
\langle 
\mathcal{T}_n(-L)  \overline{\mathcal{T}}^2_n(-\ell)  \mathcal{T}^2_n(0)  \overline\mathcal{T}_n(L) 
\rangle_\beta
=\nonumber\\ 
= c_n \, c_n^{(2)}
\left[\frac{\beta}{\pi } \,\sinh \left(\frac{\pi \ell}{\beta}\right)\right]^{-2\Delta^{(2)}_n}
\left[\frac{\beta}{\pi } \,\sinh \left(\frac{2\pi L}{\beta}\right)\right]^{-2\Delta_n }
\frac{\mathcal{F}_n(x)}{x^{\Delta^{(2)}_n}}\,.
\label{interme2}
\eea
Notice that, taking $L\to\infty$ for finite integer $n$, the above expression vanishes, signaling that we 
should first take the replica limit and only after the large $L$ one, as anticipated above. 

Thus, taking the replica limit at finite $L$, we find 
\be
\label{corr beta L finite}
\lim_{n_e \rightarrow 1} \ln
\big[\langle 
\mathcal{T}_n(-L)  \overline{\mathcal{T}}^2_n(-\ell)  \mathcal{T}^2_n(0)  \overline{\mathcal{T}}_n(L) 
\rangle_\beta\big]
 = 
\mathcal{E}_{\rm naive} +\frac{c}{4} \ln (x) + f(x),
\ee
where $\mathcal{E}_{\rm naive}$ is given by Eq. (\ref{naive}) and we defined
\be
f(x) \equiv \lim_{n_e \rightarrow 1} \ln [\mathcal{F}_{n_e}(x)].
\ee
In the above expression, the $L$ dependence is fully encoded in the harmonic ratio $x$,
thus we can  take the limit $L \to +\infty$, obtaining the logarithmic negativity (restoring the $a$ dependence)
\be\fl
\label{neg T}
\mathcal{E} = 
\mathcal{E}_{\rm naive} -\frac{\pi c \ell}{2\beta} + f(e^{-2\pi \ell/\beta}) =
\frac{c}{2} \ln \Big[\frac{\beta}{\pi a} \sinh \Big(\frac{\pi \ell}{\beta}\Big)\Big]
 -\frac{\pi c \ell}{2\beta} + f(e^{-2\pi \ell/\beta}) +2\ln c_{1/2} \,.\;
\ee
This formula represents the main result of our paper and now we are going to discuss all its implications, which are numerous.
The most remarkable and important feature of Eq. (\ref{neg T}) is that the linear term in $\ell/\beta$ exactly cancels 
the large temperature divergence of ${\cal E}_{\rm naive}$ which was the main motivation of this calculation. 
For large temperature and at fixed $\ell$ Eq. (\ref{neg T}) is valid 
only as long as $\beta\gg a$ and so it never becomes negative, as it could seem at a first 
superficial look.
Another fundamental feature is that 
the logarithmic negativity depends on the full operator content of the model through the function $f(x)$. 
This is a relevant difference compared to ${\cal E}_{\rm naive}$, which instead depends only on the central 
charge like the entanglement entropy. 
In the limit of very large $T$, we would expect the negativity to vanish since a quantum system should crossover to a classical one. 
However, in the above expression it is clear that this limit is dominated by some non-universal constants 
like the UV cutoff $a$ signalling that probably the vanishing at high-$T$ is not a direct feature of CFT, but it is connected 
to non-universal physics.  

Although Eq. (\ref{neg T}) depends on the UV physics via the cut-off $a$, there is a straightforward way to 
construct a universal scaling quantity by subtracting to the negativity  its value at $T=0$,  finding ($T=1/\beta$)
\be
\label{neg subtracted}
{\cal E}_s(T)\equiv
\mathcal{E}(T) - \mathcal{E}(0) 
=
\frac{c}{2} \ln \left[\frac{1}{\pi \ell T} \sinh \left({\pi \ell T}\right)\right] -\frac{\pi c  \ell T}{2} + f(e^{-2\pi \ell T}),
\ee
which depends only on the ratio $\ell/\beta=\ell T$ and it can be written as
\be
\label{neg sub lambda}
\mathcal{E}_s(T)
=
\frac{c}{2} \Big[ 
\ln(1- e^{-2\lambda}) - \ln (2\lambda)  
\Big]
 + f(e^{-2\lambda}),
 \qquad
 \lambda \equiv  {\pi \ell T}.
\ee

The explicit calculation of the negativity requires the evaluation of the four-point function (\ref{interme}) 
on the complex plane for some specific models, but this is a very cumbersome calculation.
Indeed this four-point function can be obtained from the known six-point functions of twist fields ${\cal T}_n$ \cite{ctt-14}
and fusing two pairs of them to to ${\cal T}_n^2$, closely following Ref. \cite{us-long}.
However,  this procedure is very cumbersome even for the easiest CFTs.
Furthermore, even if we would manage to calculate this four-point function, the replica limit is an
even more difficult task. In fact, none of the explicitly calculated four-point functions
has been continued to arbitrary complex $n$ in order to obtain a closed form of the negativity 
(see Ref. \cite{us-long} for a discussion of this point).
Anyhow, Eq. (\ref{neg T}) contains most of the physics of the negativity, even if we do not know 
the explicit form of $f(x)$. Indeed, the most interesting limits $\ell\gg \beta$ and $\beta\gg \ell$
can be analytically derived using the OPE 
of the twist fields (or better the short-interval expansion) developed in Ref. \cite{cct-11}.

\subsection{High  temperature regime}

The high temperature regime is given by $\beta \ll \ell< L$ or, equivalently, $x\to 0$. 
We wrote Eqs. (\ref{interme2}) and (\ref{neg T}) in such a way that this limit is straightforward. 
Indeed according to Eq. (\ref{bc on F}), the scaling functions ${\cal F}_n(x)$ tends to a constant value 
independent of $x$ and so will do its replica limit $f(x)$.
Consequently for $\ell\gg \beta$, the negativity (\ref{neg T}) attains a constant non-universal value because,
as already stressed above, the leading linear terms in $\ell$ cancel. 
The study of the approach to this constant value requires the knowledge of the 
subleading terms in the OPE (\ref{OPE 2}), which is a more cumbersome and involved calculation. 
We limit ourselves to mention that since the replica limit goes outside the realm of unitary CFT 
(as signalled by the fact that the scaling dimension $\Delta^{(2)}_{n_e}$ tends to the negative value $-c/4$), 
logarithmic subleading corrections to this asymptotic behaviour could also arise (as indeed happens in the case of 
other short-distance expansions \cite{us-long}).

\subsection{Low  temperature regime}

The low temperature regime is attained for $\beta \gg \ell$, i.e. when the length of the interval $A$
is much smaller than the inverse temperature. 
It is however important to stress that a reasonable result is obtained only for $L\gg\beta$, and so, in order to recover 
the zero temperature limit, we should first take the limit $L\to\infty$ and only after send $\beta\to\infty$. 
In this limit we have $x=e^{-2\pi \ell/\beta}\to 1$, so Eq. (\ref{neg T})
gives the correct zero temperature result (\ref{neg2pt}), because the normalisation $\mathcal{F}_n(1)=1$ 
implies $f(1)=0$.

The correction to the zero temperature behaviour are more involved but they can be obtained as an 
application of the short-length expansion developed in \cite{cct-11}, which requires to take 
into account the next order in the OPE (\ref{OPE id}).
The short-length expansion can be thought as the following OPE (for a more precise treatment we refer 
to \cite{cct-11})
\be\fl 
\mathcal{T}_n(u)  \overline{\mathcal{T}}_n(v)  =
\sum_{\{k_j\}} 
\frac{c_n}{|u-v|^{2\Delta_n-\sum_j(\Delta_{k_j}+\overline{\Delta}_{k_j})}}
d_{\{k_j\}}\prod_{j=1}^n\phi_{k_j}(u_j)\,,
\qquad u\to v,
\ee
where $\{\phi_k\}$ denotes a complete set of local fields for a
single copy of the CFT, $u_{j}$ is the point $u$ on the $j$-th sheet, and
$d_{\{k_j\}}$ are dimensionless universal numbers, 
which for primary operators can be explicitly calculated as \cite{cct-11}
\be
d_{\{k_j\}}^{(n)}=n^{-\sum_j(\Delta_{k_j}+\overline\Delta_{k_j})}
\langle\prod_{j=1}^n\phi_{k_j}\big(e^{2\pi ij/n}\big)\rangle_{\mathbb C}\,.
\label{main}
\ee
The leading term as $u\to v$ is given by taking all the $\Delta_{k_j}=0$, that is $\phi_{k_j}={\mathbb I}$, the identity operator, recovering 
Eq. (\ref{OPE id}).
The next term comes from taking all the $k_j=0$ except one. 
However in this case $\phi_k$ cannot be primary, since the one-point function $\langle\phi_k\rangle_{\mathbb C}$ 
in Eq. (\ref{main}) would vanish.
The most interesting case is when $\phi_{k_j}$ is a component of the stress tensor on the sheet $j$, $T_j$ or $\overline T_j$ which 
has been studied in \cite{cct-11} leading to $d_T^{(n)}=d_{\overline T}^{(n)} =1/6(1-1/n^2)$. 
The next contribution comes from taking two of the $k_j$ to be non-zero. 
Since the product must couple to the identity block this requires the two operators to be in the same block, and if they
are real and primary they must be the same operator. 
Thus, taking $k_{j_1}=k_{j_2}=k$ and $j_1\neq j_2$, we have \cite{cct-11}
\be
d_{k}^{(j_1j_2)}=\frac{n^{-2(\Delta_k+\Db_k)}}{\big[e^{2\pi ij_1/n}-e^{2\pi ij_2/n}\big]^{2\Delta_k}
\big[e^{-2\pi ij_1/n}-e^{-2\pi ij_2/n}\big]^{2\Db_k}}\label{eq:dkjj}\,.
\ee

For the low temperature expansion of the negativity we need to consider the OPE of ${\cal T}_n^2$, which can be 
easily written in terms of the ones of ${\cal T}_n$ since, for $u\to v$, we have
\be
\mathcal{T}^2_n(u)  \overline{\mathcal{T}}^2_n(v)  =
\left\{ \begin{array}{ll}
\displaystyle
\mathcal{T}_n(u)  \overline{\mathcal{T}}_n(v) 
\hspace{.5cm}& 
\textrm{odd $n$}
\\
\displaystyle
(\mathcal{T}_{n/2}(u)  \overline{\mathcal{T}}_{n/2}(v) )^2
\hspace{.5cm}& 
\textrm{even $n$}
\end{array}
\right. .
\ee
Let us now focus on the even case, when for $\ell\to 0$ we have
\bea\fl
\mathcal{T}^2_{n_e}(0)  \overline{\mathcal{T}}^2_{n_e}(\ell)= \frac{c_{n_e/2}^2}{\ell^{2\Delta_{n_e}^{(2)}}}
\sum_{k_j} \ell^{\sum_j (\Delta_{k_j}+\overline{\Delta}_{k_j})} d_{\{k_j\}}^{(n_e/2)}\prod_{j=1}^{n_e/2}\phi_{k_j}(\ell_j)
\nonumber\\ \fl \hspace{6cm}\times
\sum_{k_l} \ell^{\sum_l (\Delta_{k_l}+\overline{\Delta}_{k_l})} d_{\{k_l\}}^{(n_e/2)}\prod_{l=1}^{n_e/2}\phi_{k_l}(\ell_l).
\label{sum2}
\eea
The leading term in this expression is again obtained by taking all the $\Delta_{k_j}=\Delta_{k_l}=0$, 
that is $\phi_{k_j}=\phi_{k_l}={\mathbb I}$.
A first non-trivial contribution is obtained by choosing the identity in one of the two sums and one $k_j\neq 0$ in the other.
This contribution comes with a factor $2$, one for each sum. 
Let us consider explicitly the contribution of the stress-energy tensor $T_j$ of the sheet $j$ for which we have
\be
\mathcal{T}^2_{n_e}(0)  \overline{\mathcal{T}}^2_{n_e}(\ell)= \frac{c_{n_e/2}^2}{\ell^{2\Delta_{n_e}^{(2)}}}
\left(1+\dots+ 2\ell^2 d_T^{(n_e/2)} T_j(\ell)  +
\dots\right)\,.
\ee
Let us recall that the stress energy tensor can be inserted in $n_e/2$ possible sheets and that one should take also into 
account the antiholomorphic component $\overline T_j$, meaning that the above contribution should be multiplied by $n_e$. 
Thus inserting this expression directly in Eq. (\ref{4point cylinder}) we obtain 
\bea\fl
\langle 
\mathcal{T}_{n_e}(-L)  \overline{\mathcal{T}}^2_{n_e}(-\ell)  \mathcal{T}^2_{n_e}(0)  \overline{\mathcal{T}}_{n_e}(L) \rangle_\beta=
\\ \hspace{1cm}
= \frac{c_{n_e/2}^2}{\ell^{2\Delta_{n_e}^{(2)}}} \left \langle 
\mathcal{T}_{n_e}(-L) \left(1+\dots+ 2 n_e \ell^2 d_T^{(n_e/2)} T_j(\ell) +\dots  \right)  \overline{\mathcal{T}}_{n_e}(L) \right\rangle_\beta\,.
\nonumber
\label{ope cylT}
\eea
We are interested in the replica limit $n_e\to 1$ of this expression. In this case, since the operator is inserted on a single sheet, 
we have $\lim_{n\to 1} \langle {\cal T}_n(-L) O_j (x) \overline{\cal T}_n (L)\rangle_{\beta}= \langle  O (x)\rangle_{\beta}$,
 so that 
\bea\fl
\lim_{n_e\to 1}\langle 
\mathcal{T}_{n_e}(-L)  \overline{\mathcal{T}}^2_{n_e}(-\ell)  \mathcal{T}^2_{n_e}(0)  \overline{\mathcal{T}}_{n_e}(L) \rangle_\beta&=&
\frac{c_{1/2}^2}{\ell^{2\Delta_{1}^{(2)}}} \left\langle 
\left(1+\dots+ 2\ell^2 d_T^{(1/2)} T(\ell)  +\dots \right)  \right\rangle_\beta=
\nonumber \\\fl &=& 
c_{1/2}^2{\ell^{c/2}}  \left(1 +\dots+ 2\ell^2 \frac{1}2 \frac{\pi^2 c}{12 \beta^2}+\dots \right)
\,,
\eea
where in the second line we used $\langle T\rangle_{\beta} =- \pi^2 c/(6 \beta^2)$, $d_T^{(1/2)}=-1/2$ and $\Delta_{1}^{(2)}=-c/4$. 
Taking the logarithm we have the negativity
\be
 {\cal E}=\frac{c}2\ln \ell+2\ln c_{1/2}+\dots +c \frac{\pi^2 \ell^2}{12 \beta^2}+\dots \,.
\ee
Notice that this expression coincides with low temperature expansion of ${\cal E}_{\rm naive}$,
which indeed should be fully understood in terms of insertions of the stress energy tensor.

Primary operators contribute, for example, by choosing the identity 
in one sum in Eq. (\ref{sum2}) and taking two of the $k_{j_1}=k_{j_2}=k$ to be non-zero in the other, 
since the insertion of a single primary has zero prefactor as in Eq. (\ref{main}). 
Thus, we have (assuming also for simplicity  $\Delta_{k}=\overline \Delta_{k}$)
\bea\fl
\langle 
\mathcal{T}_{n_e}(-L)  \overline{\mathcal{T}}^2_{n_e}(-\ell)  \mathcal{T}^2_{n_e}(0)  \overline{\mathcal{T}}_{n_e}(L) \rangle_\beta=
\\ \fl \hspace{1cm}
= \frac{c_{n_e/2}^2}{\ell^{2\Delta_{n_e}^{(2)}}}  \left\langle 
\mathcal{T}_{n_e}(-L) \left(1+\dots+ 2 \ell^{4\Delta_k} d_k^{(j_1,j_2)} \phi_{k}(\ell_{j_1})\phi_{k}(\ell_{j_2}) +\dots  \right)  \overline{\mathcal{T}}_{n_e}(L) \right\rangle_\beta\,.
\nonumber
\label{ope cylTprim}
\eea
In this case, the calculation of the constant in front of $\ell^{4\Delta_k}$ is more cumbersome (because we
cannot take averages on different sheets separately) and also the replica limit is not as straightforward as in the previous case.
However, from the above expression, we immediately conclude that the contribution to the negativity is of the form
\be
 {\cal E}=\frac{c}2\ln \ell+2\ln c_{1/2}+\dots+ C_{k} \Big(\frac{\ell}{\beta}\Big)^{4\Delta_k}+\dots \,,
\ee
where the constant $C_k$ is still undetermined, but it is reasonable that, since it comes from the limit $n_e\to1$, 
for some operators will be negative (in the same way $d_T^{(n)}$ is negative for $n=1/2$).

If we consider the small temperature expansion of ${\cal E}_s(T)$ in Eq. (\ref{neg subtracted}) we  have
\be
{\cal E}_s(T)={\cal E}(T)-{\cal E}(T=0)= 
c\frac{(\pi \ell T)^2}{12} + C_k (\ell T)^{4\Delta_k} +\dots \,.
\ee
In this expression there are two different contributions: one analytic 
from the stress energy tensor (which coincide with the expansion of ${\cal E}_{\rm naive}$)
and a non-analytic and non-trivial $4\Delta_k$ where $\Delta_k$ is the scaling dimension 
of the most relevant operator coupling with $\mathcal{T}_n^2$.
Which contribution is more relevant depends on the precise operator content of the theory and 
indeed it will be interesting to provide specific examples which will be considered in following publications. 
At higher-order the computation of negativity is very complicated because it gets contributions 
from many sources (the cross terms from the sums in Eq. (\ref{sum2}), the descendent fields, etc.).


\subsection{Semi infinite systems}

A simple and important generalisation of the previous calculation is the negativity of a semi-infinite CFT on the half-line $\sigma<0$. 
In this case we have to compute the two-point function
\be
\label{2point half cylinder}
\langle 
 \mathcal{T}_n(-L)  \overline{\mathcal{T}}^2_n(-\ell) 
\rangle_\beta\,,
\ee
in the geometry where $\sigma \leqslant 0$ and the Euclidean time $\tau$ is periodic with period $\beta$.
The conformal map $z=e^{2\pi w/\beta}$ ($w=\sigma+\textrm{i} \tau$) maps the half-cylinder into the unit disk.
Using Eq. (\ref{genmap}) we have
\be
\label{2point half cylinder-disk}
\langle 
\mathcal{T}_n(-L)  \overline{\mathcal{T}}^2_n(-\ell)  
\rangle_\beta
=
\bigg(\frac{2\pi}{\beta}\bigg)^{\Delta_n+\Delta_n^{(2)}}
\frac{\langle 
\mathcal{T}_n(e^{-2\pi L/\beta})  \overline{\mathcal{T}}^2_n(e^{-2\pi \ell/\beta})  
\rangle_{\rm disk}}{
e^{2\pi \Delta_n L/\beta}\,
e^{2\pi \Delta^{(2)}_n \ell/\beta}
}\,.
\ee
On the unit disk, the two-point function of two primary operators with dimensions $\Delta_{(1)}$ and $\Delta_{(2)}$ 
  can be written as 
\be\fl 
\langle  V_1(z_1)  V_2(z_2)  \rangle_{\rm disk}
=
\frac{\mathcal{B}(y)}{(1-|z_1|^2)^{\Delta_{(1)}} (1-|z_2|^2)^{\Delta_{(2)}}},
\qquad
y = \left| \frac{z_1 -z_2}{1-z_1\bar{z}_2} \right|^2\,,
\ee
where $y$ is the harmonic ratio of the half-line and $\mathcal{B}(y)$ the universal scaling function 
with a boundary. 
In analogy to Eq. (\ref{interme}) for the plane, we can choose the scaling function of twist fields ${\cal B}_n(y)$  in such a way that 
\be\fl
\label{2point TT2 disk}
\langle 
\mathcal{T}_n(e^{-2\pi L/\beta})  \overline{\mathcal{T}}^2_n(e^{-2\pi \ell/\beta})  
\rangle_{\rm disk}
=\frac{\tilde{c}_n \tilde{c}_n^{(2)}}{(1-e^{-4\pi L/\beta})^{\Delta_n} \, (1-e^{-4\pi \ell/\beta})^{\Delta_n^{(2)}}}
\frac{{\cal B}_n(y)}{y^{\Delta_n^{(2)}/2}}\,,
\ee
where, on the half cylinder, the harmonic ratio is  
\be
y=\left(\frac{e^{-2\pi L/\beta} - e^{-2\pi \ell/\beta} }{1- e^{-2\pi (L+\ell)/\beta}}\right)^2.
\ee
In the limit $L \rightarrow +\infty$, we have $y \rightarrow e^{-4\pi \ell/\beta} $.
The non-universal constants $\tilde{c}_n$ and  $\tilde{c}_n^{(2)}$ are the boundary analogous of $c_n$
and $c_n^{(2)}$ to which are related through the Affleck-Ludwig boundary entropy \cite{al-91,cc-04,zbfs-06}. 
Plugging Eq. (\ref{2point TT2 disk}) into Eq. (\ref{2point half cylinder-disk}), we get
\bea\fl
\label{2point half cylinder-disk v2}
\langle 
\mathcal{T}_n(-L)  \overline{\mathcal{T}}^2_n(-\ell)  
\rangle_\beta
&=&
\frac{\tilde{c}_n \tilde{c}_n^{(2)} \mathcal{B}_n(y) y^{-\Delta_n^{(2)}/2}}{
[(\beta/\pi)\sinh (2\pi L/\beta)]^{\Delta_n}\,
[(\beta/\pi)\sinh (2\pi\ell/\beta)]^{\Delta_n^{(2)}} }\,.
\eea
Taking the replica limit (i.e. the limit $n_e\rightarrow 1$),  we get
\be\fl 
\label{Eneg Lfinite bdy}
\lim_{n_e\rightarrow 1} \langle \mathcal{T}_n(-L)  \overline{\mathcal{T}}^2_n(-\ell)  \rangle_\beta
 =
\frac{c}{4} \ln \left[\frac{\beta}{\pi a} \sinh \left(\frac{2\pi \ell}{\beta}\right)\right]+\frac{c}2 \ln y
 + \lim_{n_e \rightarrow 1} \ln \mathcal{B}_{n_e}(y)+ 2\ln \tilde c_{1/2},
\ee
where in particular one should notice the dependence on $\sinh 2\pi \ell/\beta$ instead of $\sinh \pi \ell/\beta$ 
obtained on the infinite line (analogously to the entanglement entropy \cite{cc-04,cc-rev}). 
We recall that the universal function $\mathcal{B}_{n}(y)$ generically depends also on the boundary state of the BCFT 
and it cannot be simply obtained from $\mathcal{F}_{n}$ in the bulk.
In the limit $L \to +\infty$  Eq. (\ref{Eneg Lfinite bdy}) gives
\be
\mathcal{E} =
\frac{c}{4} \,\ln \left[\frac{\beta}{\pi a} \,\sinh \left(\frac{2\pi \ell}{\beta}\right)\right]-\frac{\pi c\ell}{2\beta}
 +f_{\rm bdy}(e^{-4\pi \ell/\beta}) + 2\ln \tilde c_{1/2},
 \label{Ebou}
\ee
where we defined $f_{\rm bdy}(y)=  \lim_{n_e \rightarrow 1} \ln\mathcal{B}_{n_e}(y)$.
As for negativity at finite temperature in the infinite line, the linear term in $\ell$ exactly cancels the linear 
divergence of the first naive term at high temperature.

The low and high temperature limits of this expression are obtained in  analogy with the 
infinite line case and we do not repeat the calculations here. 

\subsection{Finite system}
\label{fssec}

We want to conclude this section by stressing that the calculation for finite systems, 
either with periodic boundary conditions or in a BCFT are immensely more complicated than the 
calculations reported in this paper because we should consider either a toroidal or annulus geometry. 
In these cases, the calculation of multi-point twist-field correlations is still beyond our reach.

\section{Numerical results for the harmonic chain}
\label{Sec5}

\begin{figure}[t]
\vspace{.0cm}
\begin{center}
\includegraphics[width=.48\textwidth]{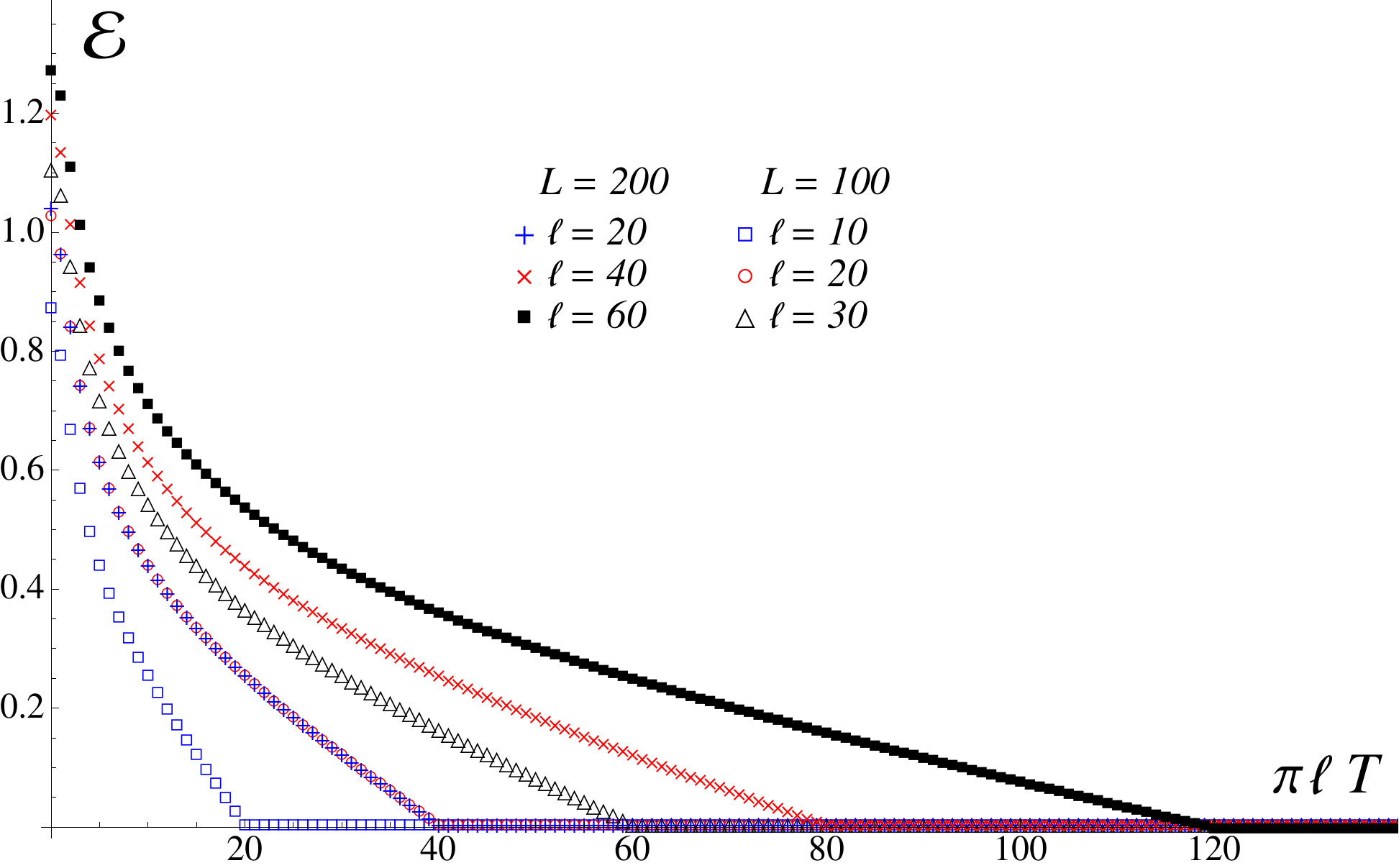}
\hspace{.3cm}
\includegraphics[width=.48\textwidth]{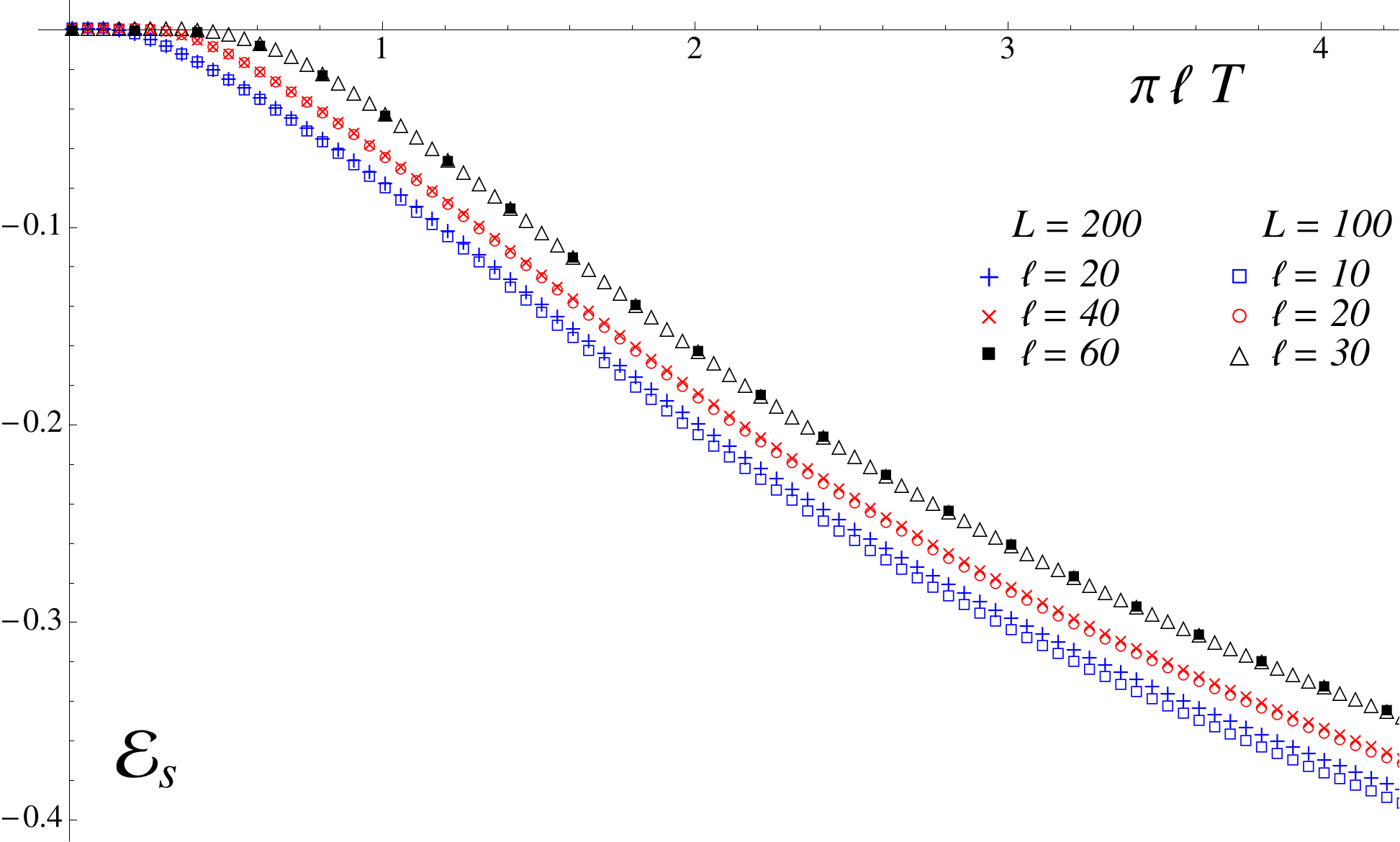}
\end{center}
\vspace{-.2cm}
\caption{
Logarithmic negativity $\mathcal{E}$ for a subsystem composed of $\ell$ contiguous sites embedded 
in a critical harmonic chain of length $L$ with Dirichlet boundary condition.  
Left: $\mathcal{E}$ for a large range of $\ell T$.
For all finite $\ell$, $\mathcal{E}$ vanishes for $T\geq T_{\rm sd}(\ell,L)$ (sudden death). 
Right: Subtracted negativity $\mathcal{E}_s \equiv \mathcal{E}(T)- \mathcal{E}(0)$ for small values of $\ell T$, i.e. 
for $T\ll T_{\rm sd}$.
}
\label{fig neg chain 1}
\end{figure}

In this section we check that the scaling forms previously obtained  are indeed recovered in numerical calculations 
of a lattice model. We consider the harmonic chain with  Hamiltonian 
\be
\label{hamiltonian HC}
H= \sum_{i=1}^L \left[
\frac{p_i^2}{2M} + \frac{M \omega^2}{2}\, q_i^2
+\frac{K}{2}(q_{i+1} - q_i)^2
\right],
\ee
where $L$ is the number of lattice sites of the chain, $M$ a mass scale, $\omega$ the one-particle oscillation frequency, 
and $K$ a nearest neighbour coupling.
The variables $p_i$ and $q_i$ satisfy standard commutation relations $[q_i,q_j] = [p_i,p_j] = 0$ and $[q_i,p_j] = i\delta_{ij}$.
We consider the harmonic chain because it is the only lattice model in which the partial transpose and the negativity can 
be obtained by means of correlation matrix techniques \cite{Audenaert02,br-04,pc-99,pedc-05}. 
The model is critical  for $\omega=0$ and its continuum limit is conformal with central charge $c=1$. 
Unfortunately, in a system with periodic boundary conditions, 
the mass term $\omega$ cannot be set equal to zero  because the $\langle q_a q_b  \rangle$ correlation function would diverge 
at zero momentum.
The presence of this zero-mode strongly affects the results also at zero temperature \cite{br-04,us-long}
and at finite $T$ gives rise to effects which obscure the universal conformal physics (indeed finite $T$ negativity 
of the harmonic chain has already been considered in Refs. \cite{AndersWinter,a-08,fcga-08}).
Thus, in order to avoid this problem and to have clean results in the conformal regime, we impose Dirichlet boundary conditions
 $q_0 = q_{L+1} =0$ and $p_0 = p_{L+1} =0$.

The Hamiltonian (\ref{hamiltonian HC}) can be diagonalised with standard techniques (see e.g. \cite{us-long,lievens}).
In order to calculate the negativity we need the dispersion relation
\be
\omega_k = \sqrt{\omega + \frac{4K}{M} \,\sin^2\left( \frac{\pi k}{2(L+1)} \right)} \,>\,\omega,
\qquad
k=1, \dots , L,
\ee
and the finite temperature correlators 
\bea \fl
\label{qq corr}
\langle q_a q_b  \rangle_\beta  
&=&
\frac{1}{L+1} \sum_{k=1}^L
\frac{1}{M \omega_k}\,
\coth\left( \frac{\omega_k}{2T}\right)
\sin\left(\frac{\pi k \, a}{L+1} \right) \sin\left(\frac{\pi k \, b}{L+1} \right)
\,\equiv\, \mathbb{Q}_{ab}\,,
\\ \fl
\label{pp corr}
\langle p_a p_b  \rangle_\beta  
&=&
\frac{1}{L+1} \sum_{k=1}^L
M \omega_k\,
\coth\left( \frac{\omega_k}{2T}\right)
\sin\left(\frac{\pi k \, a}{L+1} \right) \sin\left(\frac{\pi k \, b}{L+1} \right)
\,\equiv\, \mathbb{P}_{ab}\,.
\eea
As anticipated, in these correlators we can set $\omega=0$, while in the case of  
periodic boundary conditions the correlator $\langle q_a q_b  \rangle$ diverges when $\omega\rightarrow 0$.

From the correlators (\ref{qq corr}) and (\ref{pp corr}),  the logarithmic negativity for any subsystem $A$
of the harmonic chains can be computed following well established methods 
\cite{Audenaert02,br-04,a-08}.
Indeed the partial transposition changes the sign of all the momenta corresponding to the subsystem $A$. 
Thus we consider  the matrix $\mathbb{P}^{T_A} \equiv \mathbb{R}_A \cdot \mathbb{P}\cdot \mathbb{R}_A$, 
being $\mathbb{P}$ the matrix for the momenta (\ref{pp corr}) and  $\mathbb{R}_A$ the $L \times L$ diagonal matrix 
having $-1$ in correspondence of the sites defining $A$ and $+1$ for the remaining ones.  
Then, denoting by $\lambda_j^2$ the eigenvalues of $\mathbb{Q} \cdot \mathbb{P}^{T_A} $, 
the trace norm of the partial transpose and the negativity are given by
\be\fl 
\label{trace norm hc}
|| \rho^{T_A} || 
 = 
\prod_{j=1}^{L}
\Bigg[\,
\bigg| \lambda_j+\frac{1}{2} \bigg| - \bigg|\lambda_j -\frac{1}{2} \bigg| \,
\Bigg]^{-1}
 = 
 \prod_{j =1}^{L}
 \textrm{max}
 \bigg[ 1 ,  \frac{1}{2\lambda_j} \bigg], \quad \Rightarrow \quad
\mathcal{E}_A =\ln (|| \rho^{T_A} || )\,.
\ee
Thus the computation of $\mathcal{E}_A $ requires the diagonalisation of  a $L\times L$ matrix preventing us from  
taking the thermodynamic limit $L \rightarrow \infty$, as it is instead straightforwardly done for the entanglement entropy.

\begin{figure}[t]
\vspace{.0cm}
\begin{center}
\includegraphics[width=.8\textwidth]{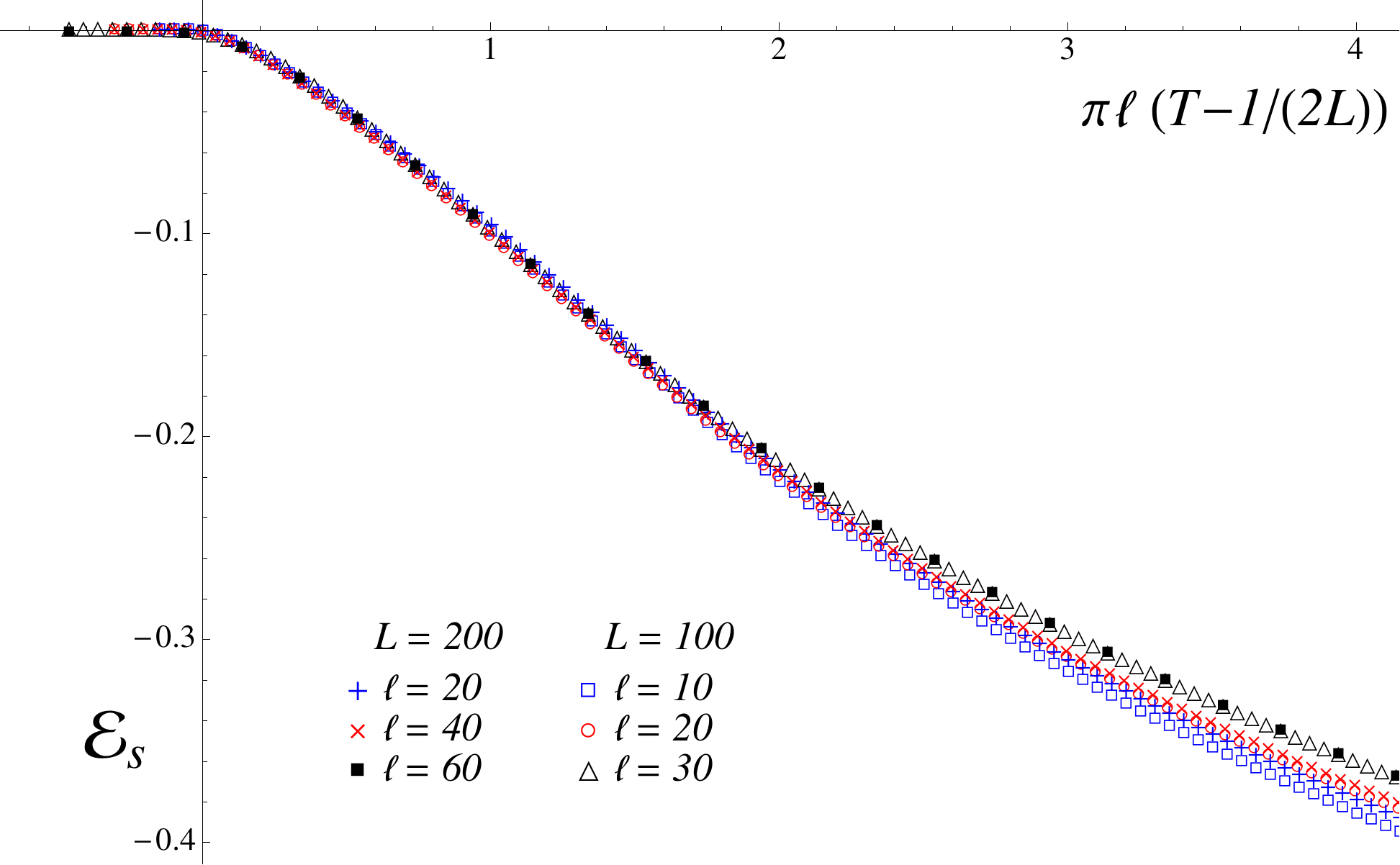}
\end{center}
\vspace{-.2cm}
\caption{
Subtracted logarithmic negativity $\mathcal{E}_s(T) \equiv \mathcal{E}(T)- \mathcal{E}(0)$ as in Fig. \ref{fig neg chain 1}. 
After shifting the temperature by $1/(2L)$, the data collapse on a master curve for temperatures 
smaller than those when finite size effects become important.
Notice that the shift $1/(2L)$ vanishes in the thermodynamic limit.
}
\label{fig neg chain 2}
\end{figure}


In the left panel of Fig. \ref{fig neg chain 1} we report the logarithmic negativity $\mathcal{E}$ at temperature $T$ for a subsystem 
composed by the first $\ell$ sites embedded in a finite chain of length $L$. 
The parameters in the Hamiltonian (\ref{hamiltonian HC}) have been fixed to $M=K=1$ and we only 
consider the critical chain with $\omega=0$. 
A first peculiar feature evident from this figure is that the logarithmic negativity becomes zero for $T\geq T_{\rm sd}$. 
This is a well known phenomenon called {\it sudden death entanglement}  (which has been first found in a 
different context \cite{ye-09}).
As already discussed for the zero temperature large distance negativity \cite{us-long}, 
the sudden death of entanglement is due to the lattice nature of the system and it is absent in a quantum field theory. 
Indeed, if one restores the lattice spacing $a$ in the above equations, it can be shown that 
$T_{\rm sd} \propto a^{-1}$ \cite{a-08}.

Thus, in order to cleanly see the CFT scaling regime, we should limit to consider a regime of temperature 
in which the lattice effects did not take place yet, i.e. $T\ll a^{-1}$. 
A zoom for these value of temperatures is shown in the right panel of Fig. \ref{fig neg chain 1}. 
In this figure we also subtracted to the negativity its zero temperature value, i.e. we consider the quantity 
${\cal E}_s(T)$ in Eq. (\ref{neg subtracted}), which, 
from CFT, we expect to be a function of the scaling variable $\pi \ell T$.
Thus all data at different $\ell$ and $T$ should collapse on the same curve given by Eq. (\ref{Ebou}), i.e.  
\be
\mathcal{E}_s(T) =
\frac{c}{4} \,\ln \frac{\sinh (2\pi \ell T)}{\pi \ell T}-c\frac{\pi \ell T}{2}
 +f_{\rm bdy}(e^{-4\pi \ell T}).
\ee
This is however not the case for the data in the right panel of  Fig. \ref{fig neg chain 1}.

The origin of this discrepancy is easily understood and can be traced back to the fact that we are studying finite chains of length $L$.
Although, we always consider the regime $\ell\ll L$, the temperature is slowly increased starting from $0$.
Therefore, there is an interval of temperatures in which $\beta > L$ and so the system behaves such as it was effectively at 
zero temperature since the finiteness of the imaginary time direction is less important than the finiteness in real space. 
And in fact, it is rather evident from the right panel of Fig. \ref{fig neg chain 1} that the negativity is practically constant for 
very small temperatures.
At a closer look, it should be also clear that $\mathcal{E}_s$ starts decreasing when $T> 1/L$.
In order to exactly take into account this finite size effect within CFT, we should consider the theory on an 
annulus, which, as stressed in Sec. \ref{fssec}, is still beyond our possibilities. 
However, we can quantitatively  describe this phenomenon as follows. 
We can introduce a threshold temperature $T_L$ which coincides with the point 
where the negativity moves from its zero temperature value.
We can then consider the negativity as function of $T-T_L$.
It is evident from the figure that $T_L\propto L^{-1}$, but we can also exactly guess the pre-factor from the 
following heuristic reasoning. 
On the annulus, we expect the negativity to be a function of 
$2\pi \ell T$ (as in Eq. (\ref{Ebou}) for infinite system)  and of $\pi\ell/L$ (as in finite size at $T=0$).
Thus $T_L$ should be $ T_L =1/(2L)$.
The results for the subtracted negativity plotted as function of  $\pi \ell (T- T_L)= \pi \ell (T-1/(2L))$ are reported  
in Fig. \ref{fig neg chain 2}. 
In this final plot we see that all the curves collapse on a single universal function
for $\pi \ell T<2$ (when we expect to be in the CFT scaling regime). 
 It is important to stress that the shift $T_L= 1/(2L) \to 0$ as $L \to \infty$.

\section{Conclusions}
\label{Sec6}

In this paper we have studied the quantum field theory approach to the logarithmic negativity of a finite interval embedded in an 
infinite one dimensional system at finite temperature. 
We have focused on conformal invariant field theories. 
We have first shown that the replica limit of the twist field correlation function 
$\langle {\cal T}^2_{n} (0)\overline{\cal T}^2_{n}(\ell)\rangle_{\beta}$ in a cylindrical geometry 
provides a wrong result. 
The reason for the failure of this naive approach has been identified in the lack of decoupling 
of the sheets composing the appropriate Riemann surface, as pictorially depicted in Fig. \ref{fig_argument}.

The correct result for the negativity can be obtained by considering the four-point function of twist fields 
\be
\langle 
\mathcal{T}_n(-L)  \overline{\mathcal{T}}^2_n(-\ell)  \mathcal{T}^2_n(0)  \overline\mathcal{T}_n(L) 
\rangle_\beta\,,
\label{4concl}
\ee
where two auxiliary fields are inserted at $w=\pm L$.
From this correlation function, the negativity is obtained by first taking the replica limit and only after taking $L\to\infty$. 
In this way, we find that the logarithmic negativity of a finite interval of length $\ell$ in an infinite CFT is given by
\be\label{neg2}
\mathcal{E} = 
\frac{c}{2} \ln \left[\frac{\beta}{\pi a} \,\sinh \Big(\frac{\pi \ell}{\beta}\Big)\right]
 -\frac{\pi c\, \ell}{2\beta} + f(e^{-2\pi \ell/\beta}) +2\ln c_{1/2} ,
\ee
where $f(x)$ is a universal scaling function, depending on the full operator content of the theory, such that 
$f(0)=0$ and $f(1)={\rm const}$.
The main feature of this results are:
\begin{itemize}
\item The naive approach based on the calculation of $\langle {\cal T}^2_{n} (0)\overline{\cal T}^2_{n}(\ell)\rangle_{\beta}$
provides only the first term in Eq. (\ref{neg2}). For large temperature and at fixed $\ell$,
this first term diverges linearly in $T=\beta^{-1}$ which is clearly an unphysical result. 
This divergence is canceled by the second term in Eq. (\ref{neg2}).
\item The finite temperature negativity depends on the full operator content of the theory and not only on the central charge, 
as opposite to the finite temperature entanglement entropy.
\item The above formula suggests that the vanishing of the negativity at very high temperature is 
governed by some non-universal physics since it happens for $\beta\gg a$.
\item The quantity ${\cal E}_s(T)\equiv\mathcal{E}(T) - \mathcal{E}(0)$ is a universal function of $\ell T$ and does 
not depend on the UV physics encoded in the inverse cutoff $a$.
\item In the limit of low temperature a universal expansion can be constructed by means of the short length 
expansion developed in Ref. \cite{cct-11}. In this expansion, the stress-energy tensor contributes to first naive piece
(quite naturally since it only depends on the central charge), while the other operators are responsible of the non trivial 
function $f(x)$. 
\end{itemize}

In the case of a CFT on the semi-infinite line, we found that the negativity of an interval starting from the boundary also has the 
universal scaling form 
\be
\mathcal{E} =
\frac{c}{4} \,\ln \left[\frac{\beta}{\pi a} \,\sinh \left(\frac{2\pi \ell}{\beta}\right)\right]-\frac{\pi c\ell}{2\beta}
 +f_{\rm bdy}(e^{-4\pi \ell/\beta}) + 2\ln \tilde c_{1/2},
\ee
in which the interpretation of the various terms is analogous to the case of an infinite system. 
We have checked this last universal scaling against exact numerical computations for the critical harmonic chain.

There are a large number of open questions which surely deserve further investigation. 
The main open problem concerns the determination of the scaling function $f(e^{-2\pi \ell/\beta})$ at least for the 
simplest CFTs, but this is a very difficult task because it requires the calculation of a six-point 
function of twist fields, fusing two pairs of them, and taking the replica limit.
At the moment this looks like an impossible mission, unless one finds a way to shortcut the procedure sketched above. 
A more affordable problem is to systematically build the low temperature expansion of Eq. (\ref{4concl})
for a specific CFTs such as for the free compactified boson.
This expansion could shed some light on the general structure of the universal function ${\cal F}_n(x)$ in which 
we are interested. However, some non-perturbative  terms (in $T$) are expected to play a crucial rule (as it  
happens at $T=0$ \cite{us-letter,us-long}) in order to obtain the full scaling of Eq. (\ref{4concl}) reported Eq. (\ref{interme2}), 
in which the term $x^{-\Delta_n^{(2)}}$ is clearly non perturbative in $T$ for $x\to e^{-2\pi \ell T}$.

\section*{Acknowledgments} 
We are  grateful to Benjamin Doyon for very useful discussions and for informing us about his 
work \cite{d-14} before publication.  This work was supported by the ERC under Starting Grant  279391 EDEQS (PC).

{\it Note added}: After the completion of this work, we became aware that some of the results derived here have been 
independently found in a different way by B. Doyon and M. Hoogeveen  \cite{d-14}.

\end{document}